# Magneto-transport characteristics of $La_{1.4}Ca_{1.6}Mn_2O_7$ thin film deposited by spray pyrolysis

**P. K. Siwach[a], H. K. Singh[b] and O. N. Srivastava[a*]**

**a** Physics Department, Banaras Hindu University, Varanasi-221005, India.

**b** National Physical Laboratory, Dr K S Krishnan Road, New Delhi-110012, India.

**Abstract**

Polycrystalline thin films of double layer manganite $La_{1.4}Ca_{1.6}Mn_2O_7$ (DLCMO) have been deposited by nebulized spray pyrolysis on single crystal $LaAlO_3$ substrates. These single phase films having grain size in the range 70-100 nm exhibit ferromagnetic transition at $T_C \sim 107$ K. The short range ferromagnetic ordering due to the in plane spin coherence is evidenced to occur at a higher temperature around 225 K. Insulator/semiconductor to metal transition occurs at a lower temperature $T_P \sim 55$ K. The transport mechanism above $T_C$ is of Mott's variable range hopping type. Below $T_C$ the current-voltage characteristics show non-linear behaviour that becomes stronger with decreasing temperature. At low temperatures below $T_{CA} \sim 30$ K a magnetically frustrated spin canted state is observed. The DLCMO films exhibit reasonable low field magneto-resistance and at 77 K the magneto-resistance ratio is ~5 % at 0.6kOe and ~13 % at 3 kOe.

∗Corresponding Author: email **hepons@yahoo.com**

## Introduction

The double layer manganites belonging to the Ruddelsden-Popper series having $Sr_3Ti_2O_7$ type crystal structures have recently been studied for their colossal magneto resistance and intricate magnetic structure [1-9]. The doped layered perovskite manganites having general formula $T_{2-2x}D_{1+2x}Mn_2O_7$ (where T is a trivalent rare earth cation like La, Nd, Pr etc. and D is a divalent alkaline earth cation like Sr, Ca, Ba etc.) are a stack of ferromagnetic (FM) metal sheets composed of $MnO_2$ bilayers which are separated by the insulating $(T,D)_2O_2$ layers and thus form a natural array of ferromagnetic-insulator- ferromagnetic (FM-I-FM) junctions [1-10]. It has been observed indeed that the incorporation of the $MnO_2$ - $(T,D)_2O_2$ - $MnO_2$ junctions in the structure naturally leads to higher magneto-resistance especially the low field magneto-resistance (LFMR) at low temperatures [1, 6-10]. The individual bilayers consisting of the FM-I-FM ($MnO_2$ - $(T,D)_2O_2$ - $MnO_2$) layers are themselves weakly coupled along the c-axis resulting in a quasi two dimensional FM order in these materials and anisotropic exchange interaction. Around room temperature the double layer compounds with x=0.3, 0.4 are paramagnets and around T~270K a short range FM order due to in plane spin coherence evolves. The long range FM order corresponding to in plane and out of plane spin coherence evolves at a much lower temperature. The $La_{2-2x}Sr_{1+2x}Mn_2O_7$ compound was the first to be investigated and it has been observed to undergo a 3D FM transition at Curie temperature $T_C \approx 126K$ for doping level x=0.4 [1,2]. The epitaxial films of this compound have also been observed to exhibit anisotropy in transport properties and high field magneto-resistance in excess of 99 percent [9,10]. The Ca doped double layer compound, viz., $La_{2-2x}Ca_{1+2x}Mn_2O_7$ has been reported to exhibit higher $T_C$ values, e.g., $La_{1.6}Ca_{1.4}Mn_2O_7$ (x=0.2) has a Tc ≈ 170K which rises to 215K for a doping level x=0.25 [6-8,11]. The epitaxial thin films of the Ca doped double layer manganites have also been studied and have been observed to exhibit $T_C$ ~ 130 K and significant LFMR below $T_C$ [6-8].The observed $T_C$ in these films are lower than the corresponding bulk values.

From application point of view thin polycrystalline films having significant magneto-resistance at low magnetic field are important. The technique of spray pyrolysis has emerged as quite effective technique to grow good quality polycrystalline thin films having smaller particle size and infinite layer manganite thin films have been successfully deposited and have been found to exhibit significant magneto-resistance both at low as well as high magnetic fields [12-14]. In this paper we report the preparation, structural/micro-structural and magneto-transport characterization of $La_{1.4}Ca_{1.6}Mn_2O_7$ polycrystalline films prepared by nebulized spray pyrolysis on single crystal $LaAlO_3$ substrates.

**Experimental**

The $La_{1.4}Ca_{1.6}Mn_2O_7$ (DLCMO) films were prepared by spray pyrolysis technique employing a two-step route. Firstly, well-homogenized aqueous solution (Molarity 0.2 M) was prepared by dissolving high purity (3N) La, Ca and Mn nitrates in appropriate cationic ratio (La/Ca/Mn=1.4/1.6/2) which then sprayed on single crystal $LaAlO_3$ (LAO) substrates in 100 orientation. A nebulizer with proper modifications was used to spray the solution and during deposition the substrate temperature was maintained at ~ 300°C. After the deposition, the films were cooled to room temperature and then annealed in flowing oxygen at 950°C for 2 hours. The thickness of the films is ~700 nm. X-ray diffraction (XRD - Philips PW1710 Cu $K_{\alpha}$) and scanning electron microscope (SEM Philips XL 20) were used for structural / micro-structural characterization. AC susceptibility measurement and four-probe electrical transport measurement were carried out for magneto transport characterization.

**Results and Discussion**

The structural characterization of the DLCMO film by XRD reveals that the films are polycrystalline and single phase having a tetragonal unit cell having lattice parameters a = 3.877 Å and c = 19.254 Å. The corresponding XRD pattern is shown in figure 1. The crystallite size was also calculated employing the Scherer formula $0.9\lambda/\beta\sin\theta$ and using the full width at half maximum (FWHM) and was found to be ~ 50 nm. The surface morphology and the particle size were also evaluated by SEM. The surface morphology consists of quite densely packed nearly spherical particles of sizes distributed in the range 70-100 nm.

The magnetic transition of the DLCMO film was studied by ac susceptibility measurement. The temperature dependence of $\chi$ was measured in the temperature range 300-16 K at an ac field of $H_{ac}$ = 20 Oe and signal frequency $f_{ac}$ = 647 Hz. The measured $\chi$ - T is plotted in figure 2. As the temperature is lowered from room temperature the susceptibility starts increasing and slope of the $\chi$ - T curve changes at T~125 K. This is signature of the onset of the paramagnetic to ferromagnetic transition the midpoint of which as determined from the $d\chi/dT$ is the Curie temperature $T_C$ = 107 K. From the $\chi$ – T data it can be conjectured that at temperatures above 125 K the DLCMO film has a weak FM character due to the usual in plane or short range FM spin coherence observed in case of the double layered manganites [7]. The Mn spins and hence the $MnO_2$ planes couple strongly along the out of plane c-direction only below 125 K and a complete or long range FM ordering is established only below $T_C$ ~ 107 K. On further lowering the temperature a drop in $\chi$ - T curve is observed at $T_{CA}$ ~ 30 K suggesting the occurrence of a spin canted state. It has been observed that in layered manganites the magnetic frustration originating due to the competition between the FM double exchange involving the itinerant $e_g$ electrons and the anti-ferromagnetic (AFM) super-exchange involving localized $t_{2g}$ electrons results in the spin canted state at low temperatures [10]. In the layered manganites the inherent anisotropy of the exchange interaction seems to further modulate this effect.

The temperature dependence of the resistivity of the DLCMO film was measured by the four-probe technique in the temperature range 300-16 K and the ρ-T data is plotted in figure 3. At room temperature the resistivity is measured to be 0.278 mΩ-cm. This low value of ρ also shows that despite smaller particle size (~70 - 100 nm) and polycrystalline nature, the film is of reasonably good quality. A metallic transition is observed at $T_P$ ~ 55 K and the resistivity at the peak of the ρ - T curve is 397.5 mΩ-cm. Thus the resistivity is found to increase by more than three orders of magnitude as the temperature is lowered from room temperature to the $T_P$, the peak resistivity temperature. At a further lower temperature the ρ - T curve shows an upturn that almost coincides with the $T_{CA}$ ~ 30 K, the spin canting temperature. This shows that there is close correlation between the resistivity reversal observed at $T_R$ ~ 28 K and the magnetically frustrated spin canted state. In fact, in the spin canted regime the scattering of the conduction electrons is increased due to the canting of Mn spins and consequently the resistivity shows an upturn [10, 15, 16].

In order to further elaborate the transport mechanism, especially in the spin canted regime we have also studied the current – voltage characteristics (IVCs) of the DLCMO film in the temperature range 300 K – 5 K measured in a liquid He cryostat. The IVCs are linear in the range 300 K – $T_C$ and the non-linearity slowly builds up below $T_C$ and at $T <$ 80 K the IVCs become appreciably non linear. The IVCs taken at three different temperatures below $T_C$ are shown in figure 5 and as seen the non-linearity is especially strong below the spin caning temperature $T_{CA}$. This strong non-linear character may be attributed to the occurrence of the magnetically frustrated spin canted state that leads to an increased scattering of the conduction $e_g$ electrons and this feature has also been in the epitaxial thin films of the layer manganite by Philipp et.al. [10]. However, in case of polycrystalline films partial contributions from the grain boundaries related disorders cannot be ruled out.

In order to have an idea of the conduction mechanism in semi-conducting regime above $T_C$ where the IVCS are linear we have analyzed the temperature dependence of resistivity above the Curie temperature $T_C \sim 107$ K. In the infinite layer simple perovskite manganites the current transport in the semi-conducting regime above $T_C$ has been explained using the Mott variable range hopping of localized electrons [17]. In the present study we have found that in the temperature regime $T_C$ - 300 K the resistivity follows the Mott's variable range hopping (VRH) model expressed by the equation:

$$\rho(T) = \rho_\infty \exp\left( T_0 / T \right)^{0.25}$$

In the conventional Mott regime the parameter $T_0$ is related to the extent of the localized states and in fact the localization length $\lambda_0 \propto T_0^{-1/3}$. [18]. The plot of $\ell n(\rho)$ and $T^{-0.25}$ is shown in figure 4. The linear fit to the observed data is also shown in the figure by solid line. The value of the parameter $T_0$ corresponding to the best fit is 3.838 x $10^7$ K and under the applied magnetic field this value is found to decrease slightly.

The magneto-resistance measurements were carried out in the temperature range 300 K – 77 K and magnetic field range H = 0.0 – 3 kOe. The magnetic field dependence of the low field magneto-resistance measured at 77 K is shown in figure 6 where two distinct slopes can easily be distinguished in the MR – H curve. The first slope is rather steep and is observed at H ≤ 0.6 kOe and above this the rate of magneto-resistance increment with the field slows down but still there is no sign of saturation up to H = 3 kOe as observed in case of infinite layer manganites [12,13]. The larger value of d(MR)/dH at H ≤ 0.6 kOe suggests that the tunneling component of the magneto-resistance is dominant in this regime. The temperature dependence of the low field magneto-resistance measured at H = 1.5 kOe is shown in figure 7. The magneto-resistance is observed to decrease monotonically as the temperature is reduced and becomes less than 2 % at 185 K and finally it vanishes around 230 K. The observed field and temperature dependence of the magneto-resistance shows that the grain boundary tunneling contribution or the extrinsic component dominates at these fields and temperatures.

**Conclusions**

In the present investigation we have prepared $La_{1.4}Ca_{1.6}Mn_2O_7$ polycrystalline films by nebulized spray pyrolysis of stoichimetric aqueous solution made from respective metal nitrates. The zero field electrical and magnetic transport properties of these polycrystalline thin films have been studied in the temperature range 300 – 5 K while the magneto-resistance properties have been studied in the temperature range 300 – 77 K. A FM transition is observed at $T_C$ ~ 107 K and at $T_{CA}$ ~30 K a spin canted state is observed. In the temperature range 300-$T_C$ the current transport is through the Mott variable range hopping of electrons localized by a random magnetic potential originating due to Hund's coupling. The metal-insulator/semiconductor transition is observed at $T_P$ ~ 55K and IVCs below $T_C$ are non linear. At 77 K the low field MR is found to be ~5 % at H = 0.6 kOe and at 3 kOe the corresponding value is 13 %.

**Acknowledgements**

Financial support from the CSIR, New Delhi is thankfully acknowledged. One of the author (PKS) acknowledges CSIR, New Delhi for the award of SRF. Authors are also thankful to Professors A R Verma, C N R Rao, T V Ramakrishnan, Vikram Kumar, S B Ogale, A K Raychaudhari , Dr Kishan Lal and Dr. N. Khare for valuable discussions.

**Figure Captions:**

Figure 1. XRD profile of polycrystalline thin film of $La_{1.4}Ca_{1.6}Mn_2O_7$ (DLCMO).

Figure 2. Variation of ac susceptibility of $La_{1.4}Ca_{1.6}Mn_2O_7$ with temperature.

Figure 3. Temperature dependence of resistivity for DLCMO thin film.

Figure 4. Variation of log (resistivity) with $T^{-0.25}$ for DLCMO thin film. The open circle shows the experimental data and the solid line is the best linear fit.

Figure 5. I-V characteristics of DLCMO thin films at 65K, 28K and 5K.

Figure 6. Magneto resistance variation with applied magnetic field for DLCMO thin film at 77K.

Figure 7. Temperature dependence of MR at 1.5kG applied magnetic field.

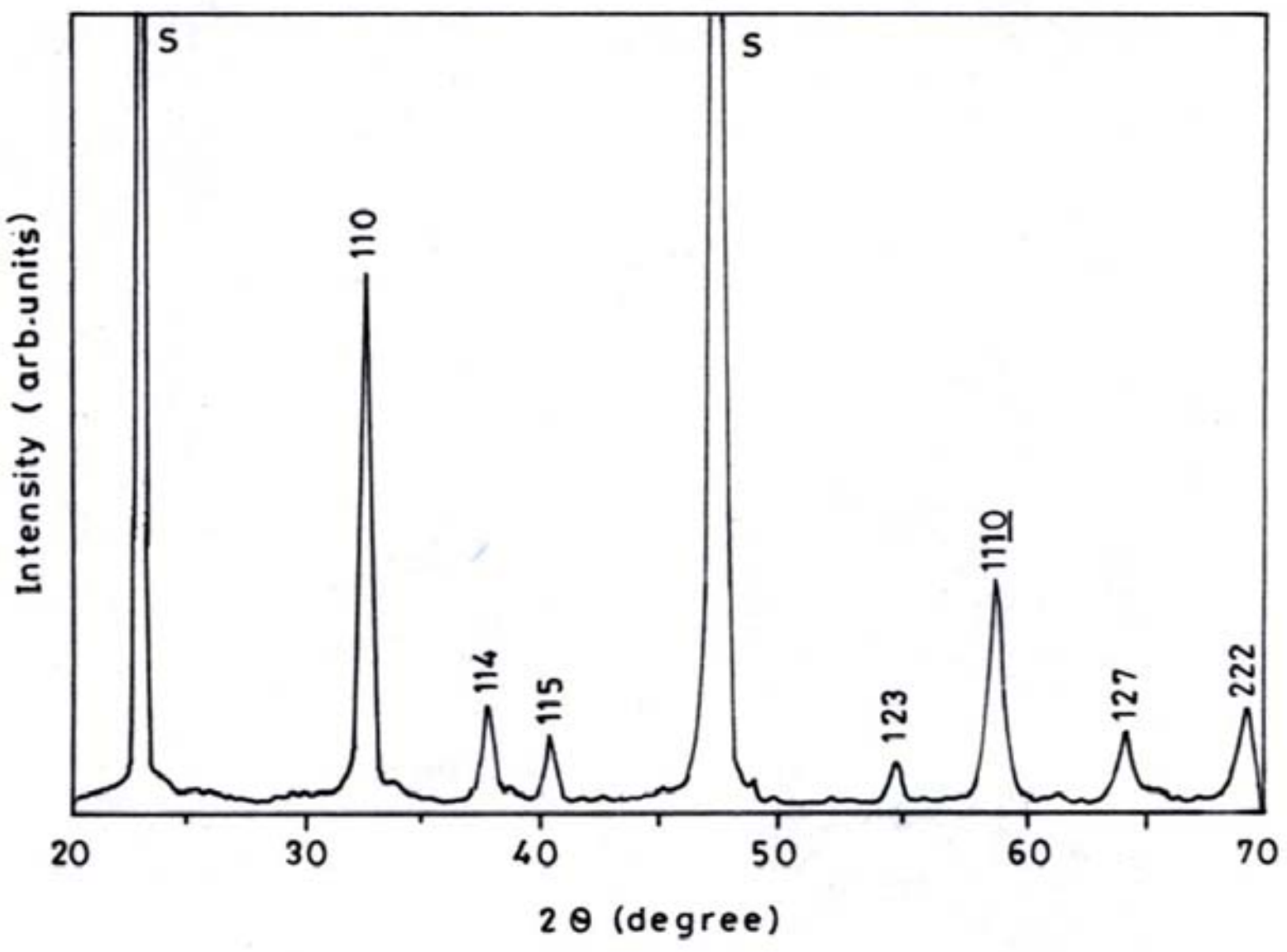

**Figure 1**

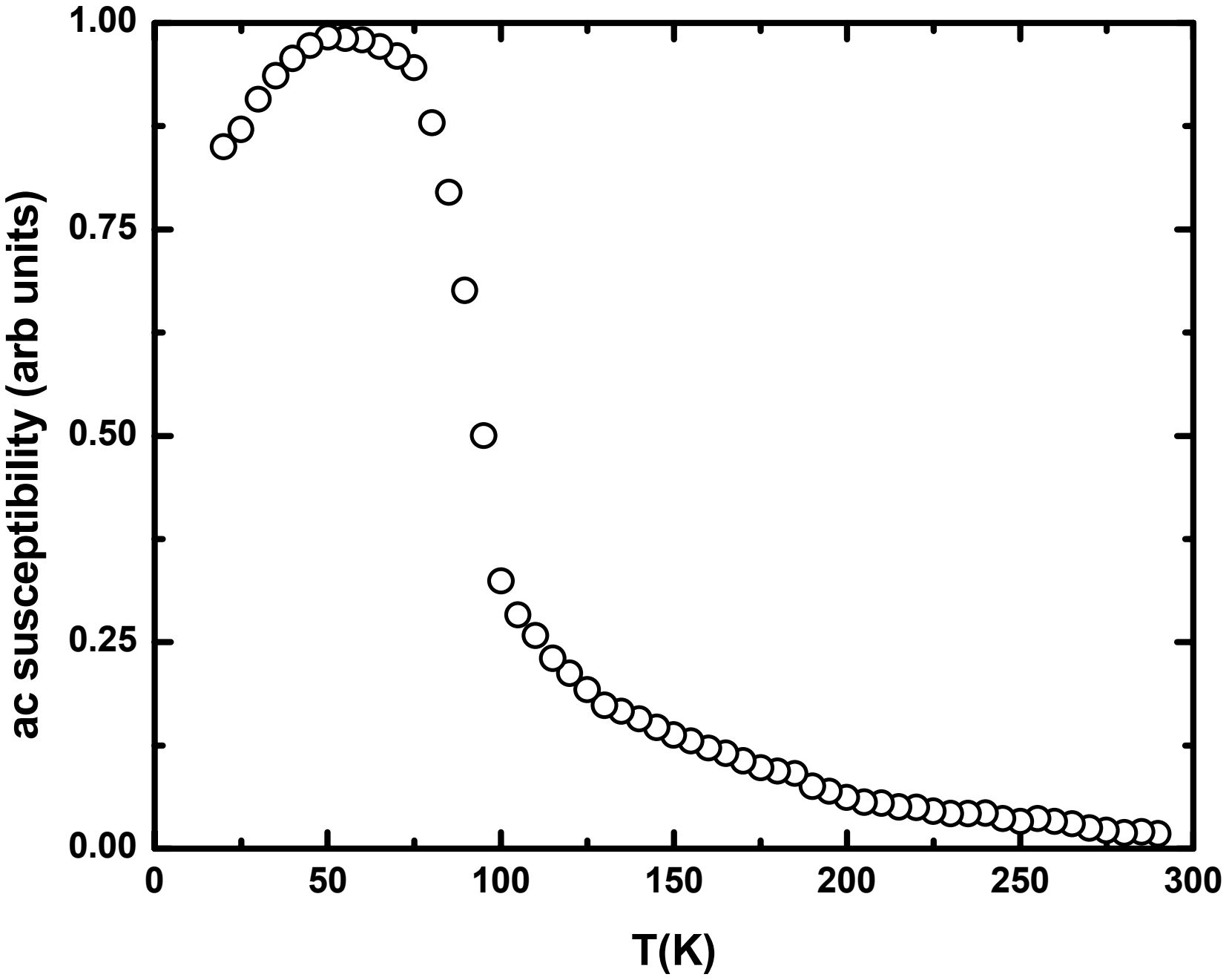

**Figure 2**

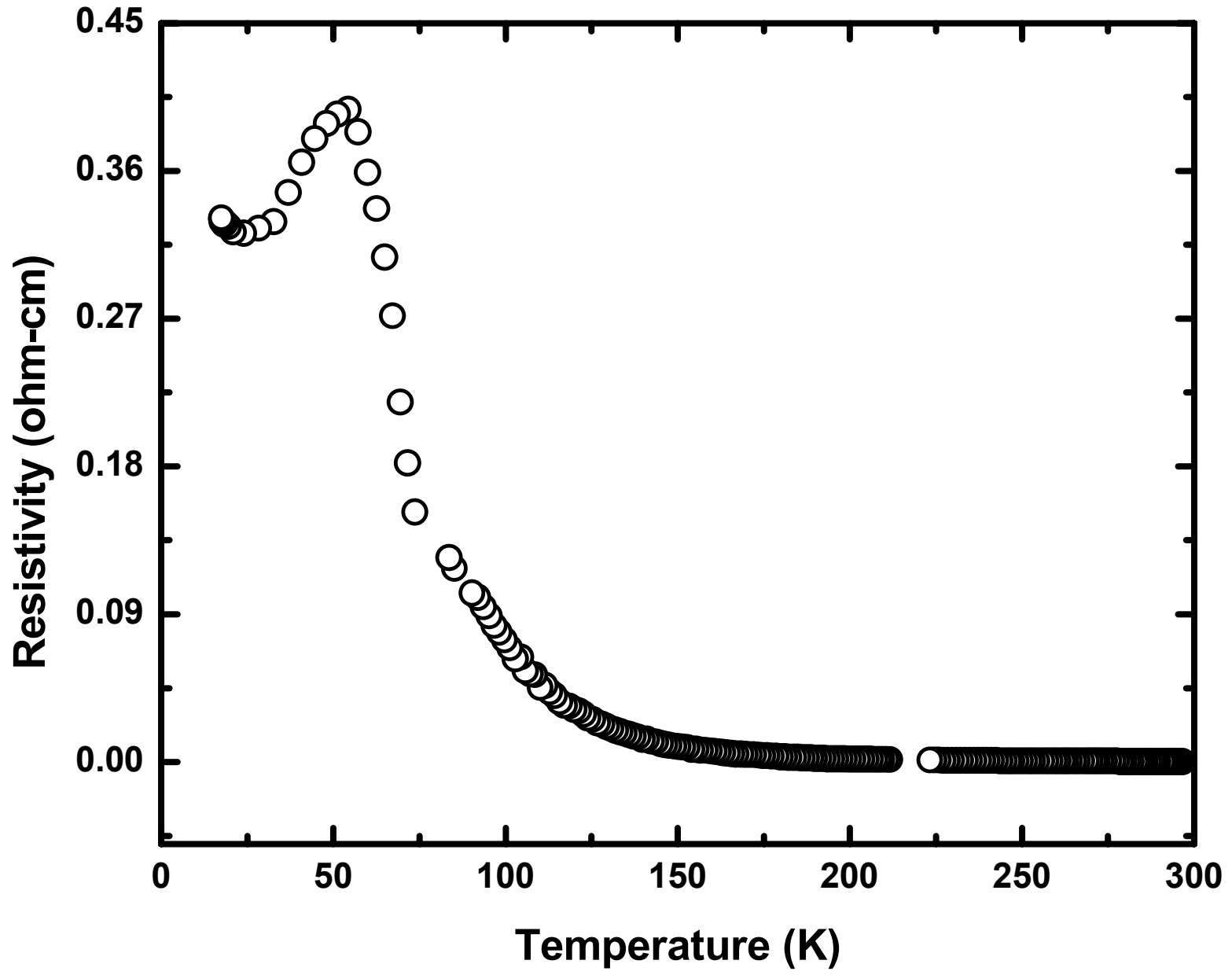

Figure 3

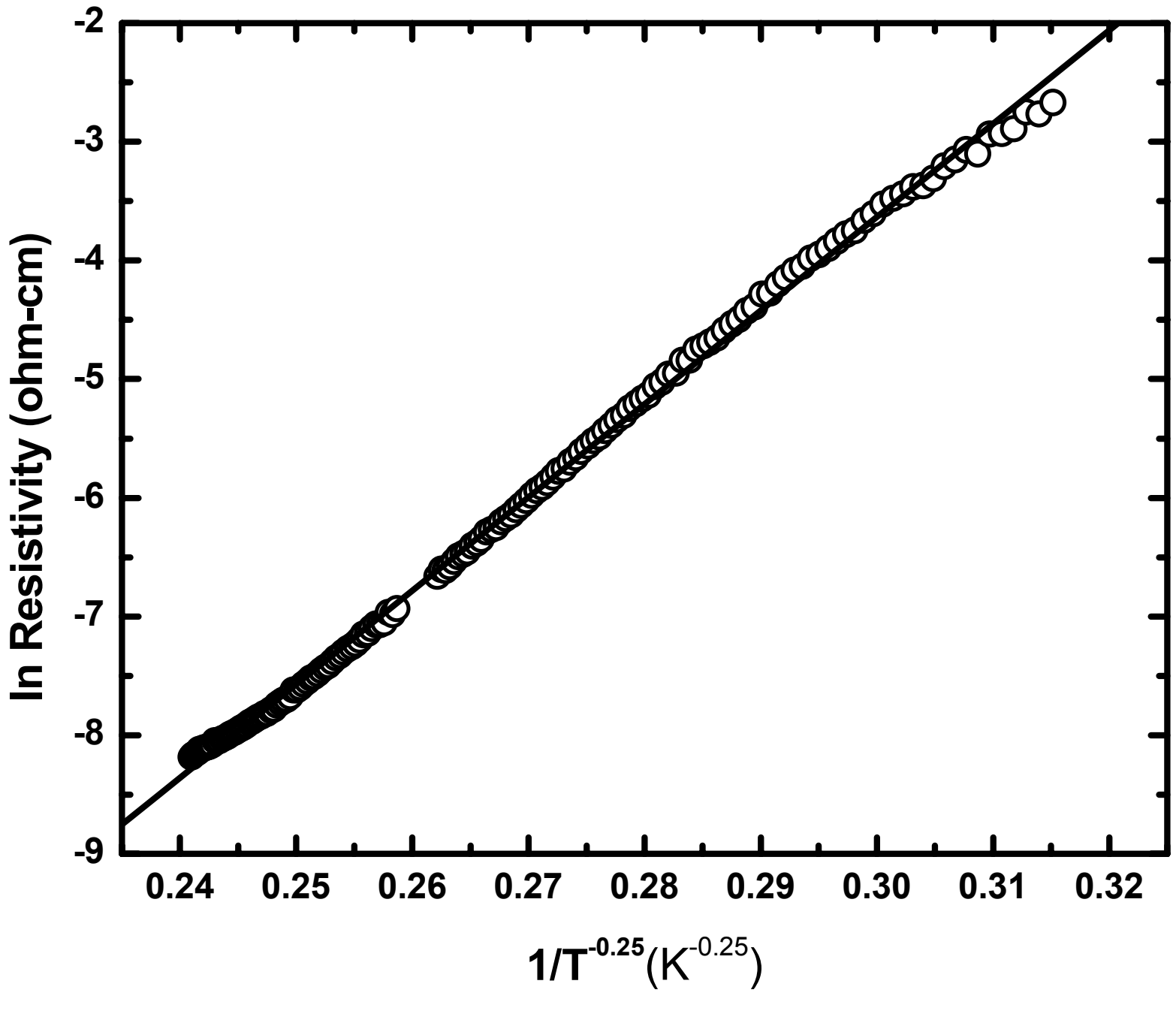

Figure 4

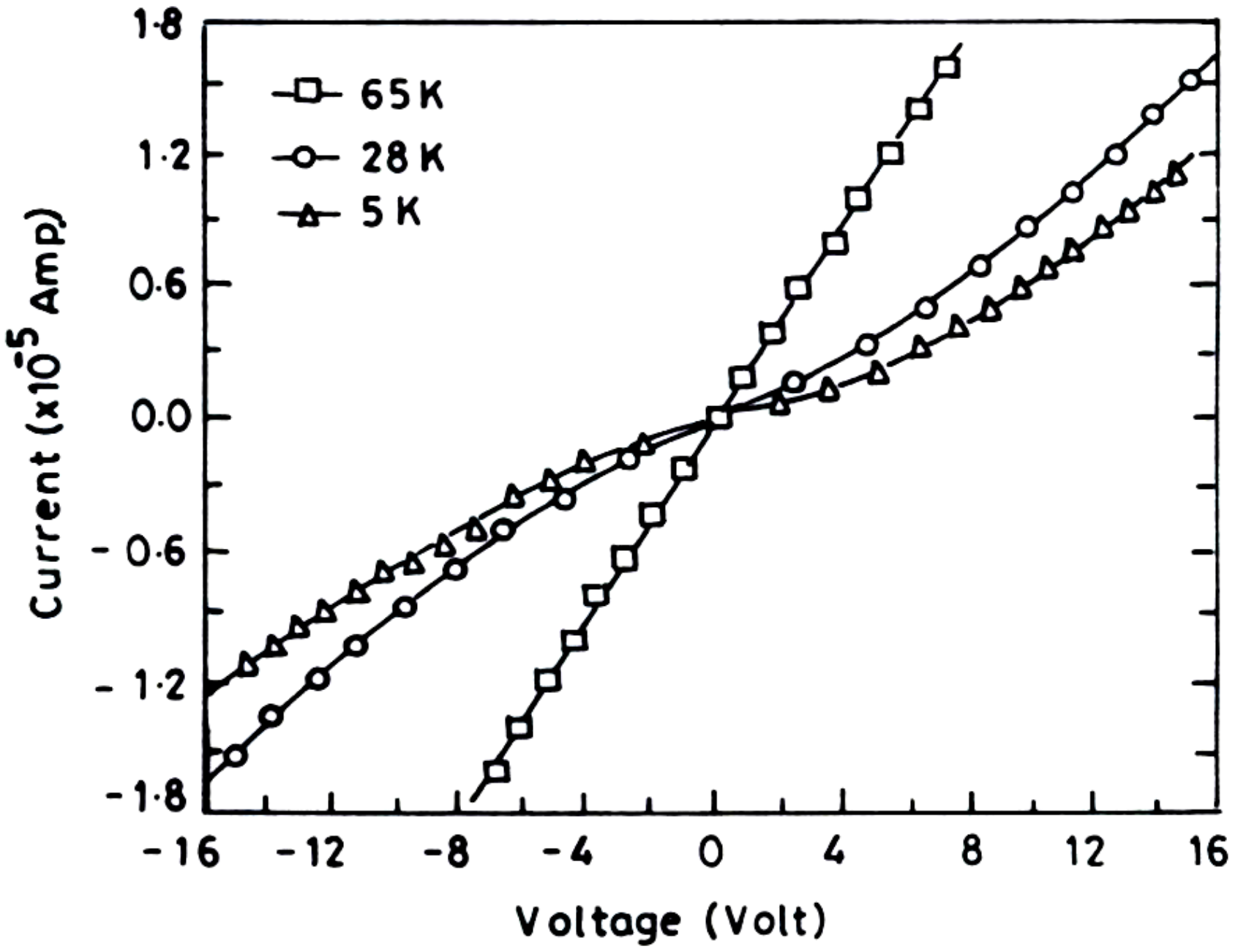

Figure 5

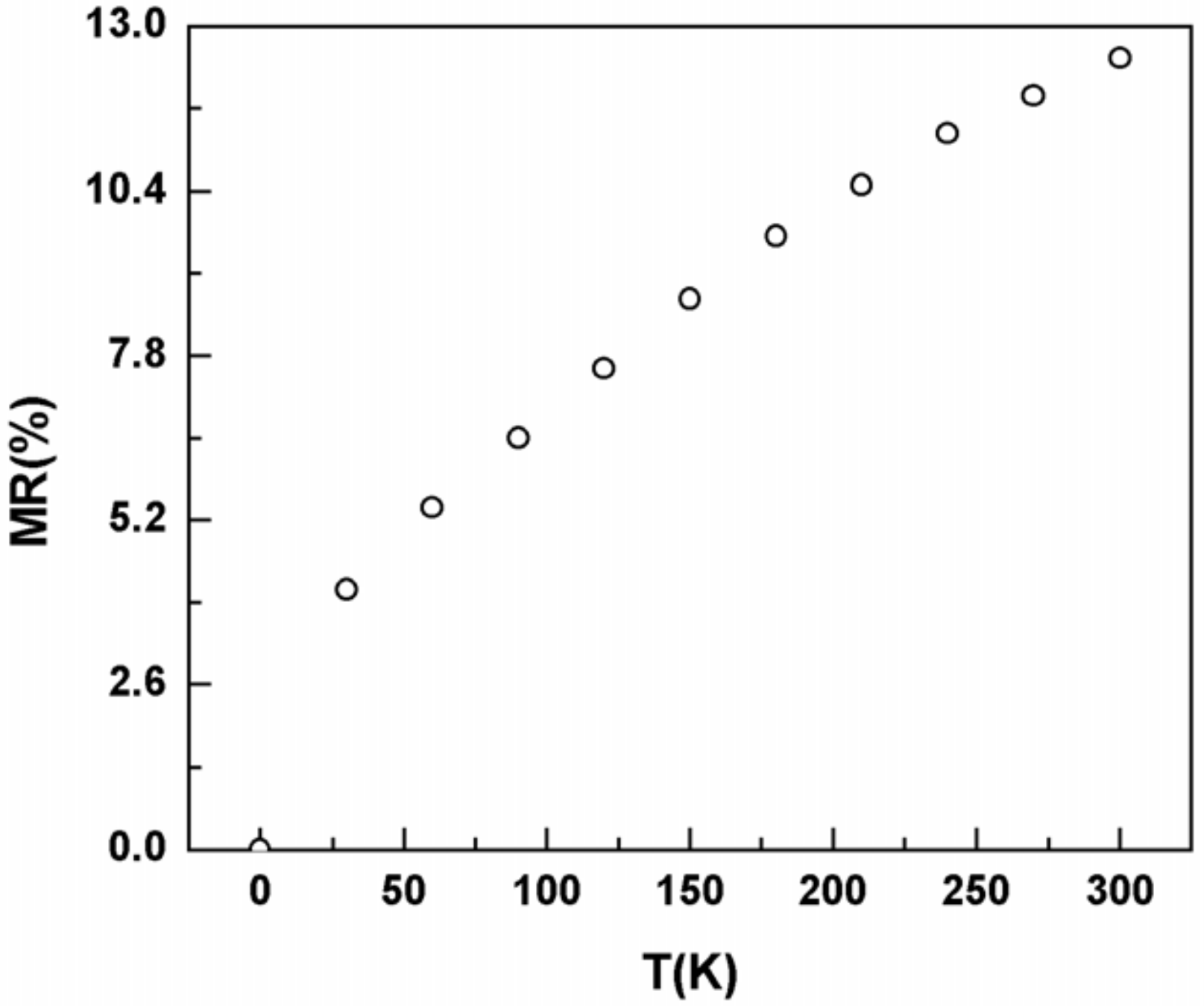

Figure 6

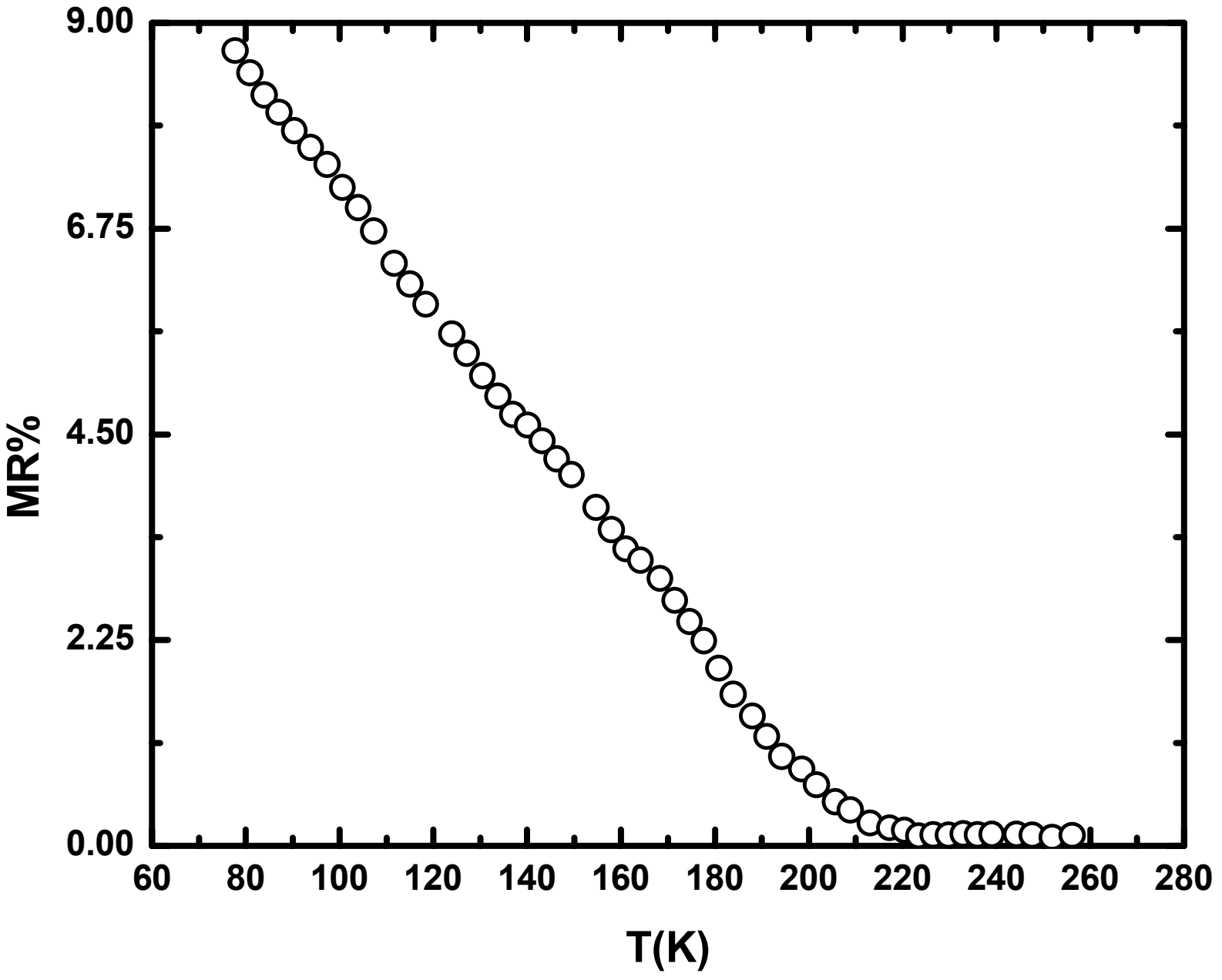

9.00
6.75
4.50
2.25
0.00
MR%
60 80 100 120 140 160 180 200 220 240 260 280
T(K)

**Figure 7.**